\documentclass[aps,prl,twocolumn,groupedaddress,floats,showpacs]{revtex4}
\usepackage{latexsym}
\usepackage{dcolumn}
\usepackage[dvips]{graphicx}
\usepackage{amssymb,amsmath,bm}
\usepackage{graphics}
\usepackage{epsf}

\begin{document}

\draft

\title{Screening, Kohn anomaly, Friedel oscillation, and RKKY
  interaction in bilayer 
graphene}

\author{E. H. Hwang and S. Das Sarma}
\address{Condensed Matter Theory Center, 
Department of Physics, University of Maryland, College Park,
Maryland  20742-4111 } 
\date{\today}

\begin{abstract}
We calculate the screening function in bilayer graphene (BLG) both in
the intrinsic (undoped) and the extrinsic (doped) regime within random
phase approximation, comparing our results with the corresponding 
single layer graphene (SLG) and the regular two dimensional electron
gas (2DEG).
We find that the Kohn anomaly is strongly enhanced in BLG. 
We also discuss the Friedel oscillation and the
RKKY interaction, which  are associated with the
non-analytic behavior of the screening function at $q=2k_F$.
We find that the Kohn anomaly, the Friedel oscillation, and the RKKY
interaction are all qualitatively different in the BLG compared with
the SLG and the 2DEG.
\end{abstract}

\maketitle
\vspace{0.5cm}



Single layer graphene (SLG), a single layer of carbon atoms
arranged in a honeycomb lattice, has 
attracted a great deal of attention, both experimentally and
theoretically, for its unusual electronic transport and the
characteristics of  relativistic  charge carriers behaving like
massless chiral Dirac fermions \cite{review}.
Bilayer graphene (BLG) consisting of two SLG  is also 
of great current interest,
both for technological
applications and  fundamental interest \cite{review,bilayer,bilayer2}.
While the band structure of SLG has a linear dispersion,
BLG has a quadratic dispersion \cite{bilayer} in the low energy regime 
making it similar to two
dimensional (2D) semiconductor systems except for the absence of a gap.  
The purpose of this work is to 
calculate the polarizability (or screening) function of
bilayer graphene within the random phase approximation (RPA).
Even though many theoretical works on Coulomb screening in SLG
have been reported \cite{Hwang_RPA,Ando,wunsch,Barlas,wang}, the analytic
investigation of Coulomb screening in 
BLG has not yet been performed.
Knowing the BLG screening function is crucial since it determines many
fundamental properties,
e.g. transport through screened 
Coulomb scattering by charged impurities \cite{transport}, Kohn anomaly in
phonon dispersion \cite{kohn},  and RKKY interaction \cite{rmp}. 
In order to understand the electronic properties
of BLG it is therefore necessary to obtain its screening function.

The BLG is in some sense intermediate between the SLG and the regular
semiconductor-based two dimensional electron gas (2DEG) since it is
chiral with a zero band gap at the Dirac point (where the electron and
the hole bands touch) similar to the SLG, but has the quadratic energy
dispersion similar to the 2DEG. For example, the $2k_F$-backscattering
is suppressed \cite{Hwang_RPA,Ando,wunsch,Barlas,wang} in SLG due to its
chiral nature whereas in the 2DEG, the $2k_F$-backscattering plays a
key role \cite{Dassarma} in determining low density and low temperature
carrier transport. We find that the $2k_F$-backscattering is restored
(and even enhanced) in the BLG because of the quadratic
dispersion and, more importantly, due to the symmetry imposed by the
two-layer structure. This qualitative difference in the screening
properties between BLG and SLG leads us to predict that transport and
other electronic properties in BLG will be qualitatively more similar
to 2DEG than to SLG in spite of the zero-gap chiral nature of BLG.

The effective BLG Hamiltonian  is
now well established in the theoretical literature. 
In the low energy regime the Hamiltonian is reduced to the $(2\times
2)$ matrix form and is given by (we use $\hbar =1$ throughout this
paper) \cite{bilayer}
\begin{eqnarray}
H_0 = -\frac{1}{2m}\left ( 
 \begin{array}{cc} 
     0        & (k_x-ik_y)^2 \\
(k_x+ik_y)^2  &      0 
 \end{array}
\right ),
\label{Ham}
\end{eqnarray}
where $m = \gamma_1/(2v_F^2)$, $\gamma_1$ is the interlayer tunneling
amplitude, and $v_F$ is the SLG Fermi velocity. 
The wave function of
Eq. (\ref{Ham}) can be written as $\psi_{s{\bf k}} =e^{i{\bf k r}}
(e^{-2i\theta_{\bf k}},s)/\sqrt{2}$ and the corresponding energy is
given by 
$\epsilon_{s{\bf k}}=sk^2/2m$, where $\theta_{\bf k} = \tan^{-1}(k_y/k_x)$
and $s=\pm1$ denote the band index. 
Using the Hamiltonian of
Eq. (\ref{Ham}) we theoretically obtain the screening  function of BLG
by calculating the polarizability and the dielectric function within
RPA.  


The static dielectric function can be written as
\begin{equation}
\epsilon(q) = 1-\frac{2\pi e^2}{\kappa q} \Pi(q),
\end{equation}
where $\kappa$ is the background dielectric constant,
and $\Pi(q)$  the polarizability.
The static BLG polarizability  is given by the bare bubble
diagram 
\begin{equation}
\Pi(q) =\frac{g}{L^2}
\sum_{{\bf k}ss'}\frac{f_{s{\bf k}}-f_{s'{\bf k}'}}
{\varepsilon_{s{\bf k}}-\varepsilon_{s'{\bf k}'}}F_{ss'}({\bf
  k},{\bf k}'),
\label{pol}
\end{equation}
where $g$ is the degeneracy factor (here $g=4$ due to valley and spin
degeneracies),  
${\bf k}'={\bf k}+{\bf
  q}$, $s,s'=\pm 1$ denote the band indices, $\varepsilon_{sk} = s k^2/2m$,
and $F_{ss'}({\bf k},{\bf k}') = (1 + ss' \cos2\theta)/2$, where
$\theta$ is the 
angle between ${\bf k}$ and ${\bf k}'$, and
$f_{sk}$ is the Fermi distribution function, 
$f_{s{\bf k}} = [\exp \{\beta(\varepsilon_{s{\bf k}}-\mu)\} + 1]^{-1}$,
with $\beta = 1/k_BT$ and  $\mu$ the chemical potential.

First, we consider 
intrinsic (i.e. undoped or ungated, with $n$ and $E_F$ both being
zero) BLG where the conduction band is empty and
the valence band  fully occupied at $T=0$.
Then we have $f_{{\bf k}+}=0$ and $f_{{\bf k}-}=1$. 
Since the conduction band is empty the polarization
is induced by the virtual interband transition of electrons from
the valence to the conduction band. The polarizability due to the
interband transition becomes 
$\Pi(q) \equiv \Pi^{0}(q)$, where
\begin{equation}
\Pi^{0}(q) =\frac{g}{2}\int \frac {d^2k}{(2\pi)^2} \left [ 
\frac{{1 - \cos2\theta}}{\varepsilon_{+{\bf k}}-\varepsilon_{-{\bf k}'}}
- \frac{{1 - \cos2\theta}}
{\varepsilon_{-{\bf k}}-\varepsilon_{+{\bf k}'}} \right ],
\label{pi0}
\end{equation}
where $\cos\theta = (|{\bf k}| + |{\bf q}|\cos\phi)/|{\bf k+q}|$.
Eq. (\ref{pi0}) can be calculated easily
\begin{equation}
\Pi^0(q) = N_0 \log4,
\label{pi0b}
\end{equation}
where $N_0=gm/2\pi$ is the BLG density of states.
Thus the intrinsic BLG polarizability is 
constant for all wave vectors. (Note that the
polarizability of ordinary 2DEG
is constant \cite{rmp} only for $q \leq 2k_F$.) The dielectric function
becomes $\epsilon(q) = 1 + q_s/q$, where
the screening wave vector is given by 
\begin{equation}
q_s = q_{TF} \log 4, 
\end{equation}
where $q_{TF}$ is the 2D
Thomas-Fermi screening wave vector \cite{rmp}, $q_{TF} = gme^2/\kappa$. 
BLG static screening is thus enhanced by a factor of $\log4$ compared
with ordinary 2D screening \cite{rmp}.
For intrinsic
BLG we can write the screened Coulomb potential as 
\begin{equation}
\phi(r) = \frac{e}{\kappa r} - \frac{e}{\kappa}\frac{\pi q_s}{2}
\left [H_0(q_s r)-N_0(q_s r) \right ],
\end{equation}
where $H_0(x)$ and $N_0(x)$ are the Struve function and the Bessel
function of the second kind, respectively. The asymptotic form at large
$r$ is $\phi(r) \sim eq_s/\kappa (q_s r)^3$. Since the screening function
is a constant for all $q$ without any singular behavior there is no
oscillatory term in the potential. This is very
different from the screening behavior of intrinsic SLG
\cite{Hwang_RPA,Ando,wunsch} or 2DEG \cite{rmp}.

For intrinsic SLG we have
\begin{equation}
\Pi^0(q) = N^{SLG}_0 \frac{\pi}{8},
\label{pi0s}
\end{equation}
where $N^{SLG}_0 = g q/(2\pi v_F)$.
The intrinsic SLG polarizability increases linearly with 
$q$, and $\epsilon(q) = 1
+ (e^2 g/\kappa v_F) (\pi/8)$, which 
gives rise to only
an enhancement of the effective background dielectric constant
$\kappa^* = \kappa  + (e^2 g/v_F) (\pi/8) $,
i.e. the screened Coulomb interaction $V(q) =
2\pi e^2/\kappa^* q$. The Coulomb
interaction in real space can be expressed by $V(r) = e^2/\kappa^* r$ for all 
$r$. Thus at large $r$ the Coulomb potential decreases as $1/r^3$ in
intrinsic BLG, but only as $1/r$ in intrinsic SLG. 


In the following we provide the zero temperature static
polarizability of extrinsic (i.e. gated or doped) BLG where 
$n,E_F \neq 0$. 
At T=0, $f_{-\bf k}=1$ and $f_{+ \bf k} = \theta(k_F-|{\bf k}|)$. Then we
can rewrite Eq. (1) as
$\Pi(q) = \Pi_{\rm intra}(q) + \Pi_{\rm inter}(q)$, where
\begin{equation}
\Pi_{\rm intra}(q) =-\frac{g}{L^2}
\sum_{{\bf k}s}\left [ \frac{f_{s{\bf k}}-f_{s{\bf k}'}} 
{\varepsilon_{s{\bf k}}-\varepsilon_{s{\bf k}'}} \right ]
\frac{1 + \cos2\theta}{2},
\end{equation}
and
\begin{equation}
\Pi_{\rm inter}(q) =-\frac{g}{L^2}
\sum_{{\bf k}s} \left [ 
\frac{f_{s{\bf k}}-f_{-s{\bf k}'}}{\varepsilon_{s{\bf
      k}}-\varepsilon_{-s{\bf k}'}}  \right ] 
 \frac{1 - \cos2\theta}{2}.
\end{equation}
$\Pi_{\rm intra}$ ($\Pi_{\rm inter}$) indicates the polarization due
to intraband (interband) transition. After angular integration over
the direction of {\bf q} we have 
\begin{eqnarray}
\Pi_{\rm intra} (q)=\frac{gm}{2\pi} \int_0^{k_F}\frac{dk}{k^3} \left
    [ k^2 - |k^2-q^2| \right . \nonumber \\
\left . + \frac{(2k^2-q^2)^2}{\sqrt{q^2-4k^2}}\theta(q-2k)
    \right ],
\end{eqnarray}
\begin{equation}
\Pi_{\rm inter} (q)=\frac{gm}{2\pi} \int_{k_F}^{\infty}\frac{dk}{k^3} \left
    [ -k^2 - |k^2-q^2| + \sqrt{4k^4+q^4}
    \right ].
\end{equation}
Then 
\begin{eqnarray}
\frac{\Pi_{\rm intra}(q)}{N_0} = \left \{ 
 \begin{array}{ll} 1-\frac{q^2}{2k_F^2}  & \mbox{if $q \leq k_F$} \\
                  \frac{q^2}{2k_F^2}-2\log \frac{q}{k_F}
                  & \mbox{if $ k_F < q < 2 k_F$} \\
          \frac{q^2}{2k_F^2}-2\log \frac{q}{k_F} - f(q) & \mbox{if $ q
            > 2 k_F$}  
\end{array} 
\right .  ,
\end{eqnarray}
\begin{eqnarray}
\frac{\Pi_{\rm inter}(q)}{N_0} =  \left \{ 
 \begin{array}{ll} -1+\frac{q^2}{2k_F^2}  + g(q) & \mbox{if $q \leq k_F$} \\
          -\frac{q^2}{2k_F^2}+2\log q + g(q) & \mbox{if $ q >  k_F$} 
\end{array} 
\right .  ,
\end{eqnarray}
with
\begin{eqnarray}
f(q) & = &
\frac{2k_F^2+q^2}{2k_F^2q}\sqrt{q^2-4k_F^2} + \log
\frac{q-\sqrt{q^2-4k_F^2}}{q+\sqrt{q^2-4k_F^2}} \nonumber \\
g(q) & = & \frac{1}{2k_F^2}\sqrt{4k_F^4+{q^4}} -\log \left [
  \frac{k_F^2+\sqrt{k_F^4+{q^4}/4}}{2k_F^2} \right ].
\end{eqnarray}
Finally, we have the extrinsic BLG static polarizability as
\begin{equation}
\Pi(q) = N_0 \left [ f(q) - g(q) \theta(q-2k_F) \right ].
\label{piq}
\end{equation}
Eq.(16) with Eq. (15) is the basic result obtained in this paper,
giving the doped BLG polarizability analytically.

\begin{figure}
\epsfysize=2.2in
\epsffile{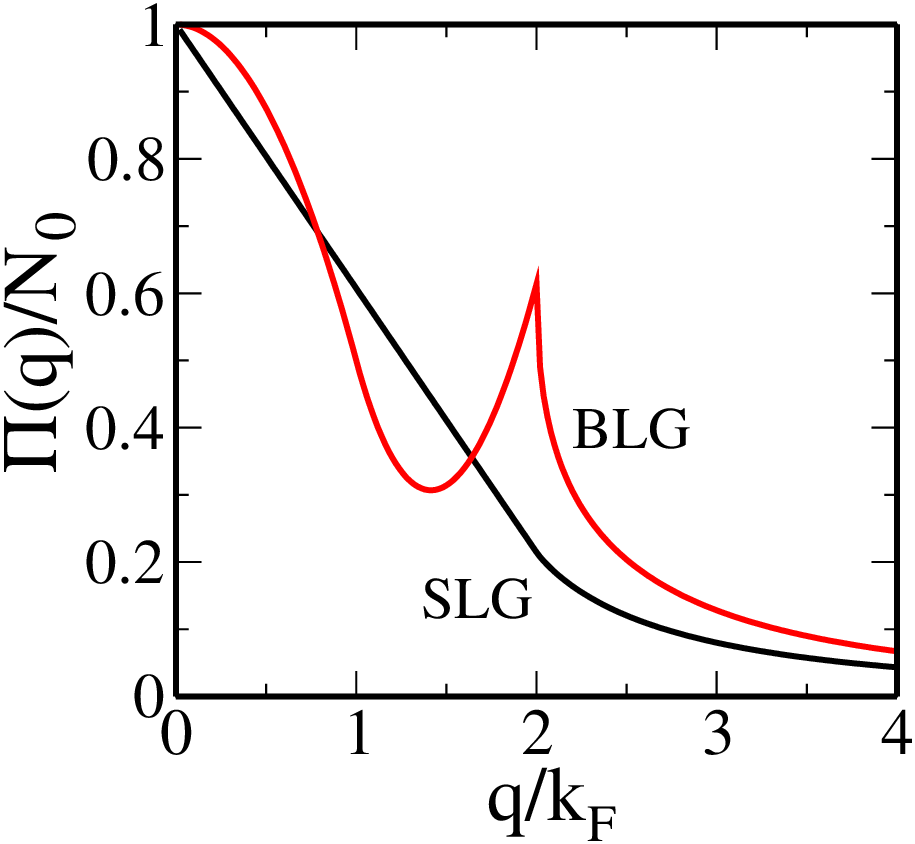}
\epsfysize=2.2in
\epsffile{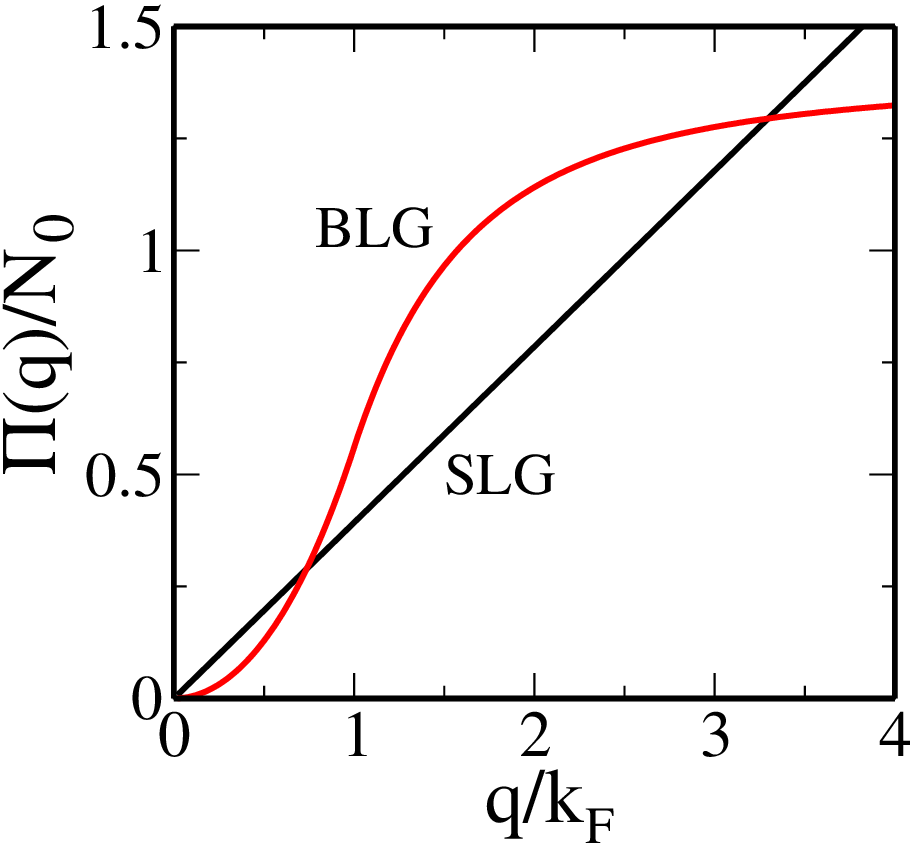}
\epsfysize=2.2in
\epsffile{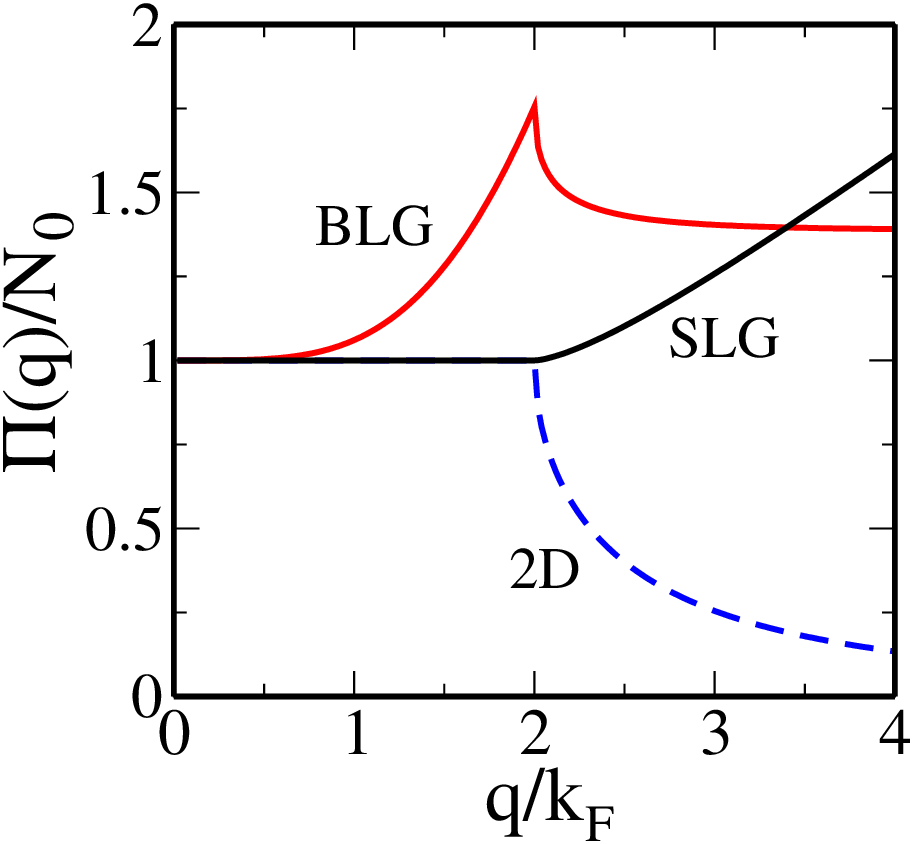}
\caption{\label{s_pol} 
(Color online) Calculated (a) intraband, (b) interband, and (c) total static
polarizability of bilayer graphene. For comparison the single layer
polarizabilities are shown. In (c) we also show the regular 2D static
polarizability (dashed line). 
}
\end{figure}

In Fig. \ref{s_pol} we show the calculated static polarizability
as a function of wave vector. 
Fig. \ref{s_pol} (a) and (b) show the calculated intraband and
interband polarizabilities, respectively, with those of single layer
graphene for comparison. Fig. \ref{s_pol} (c) shows total
polarizability of bilayer graphene.
At $q=0$ we have $\Pi_{\rm intra}(0) = N_0$ and $\Pi_{\rm inter}(0) = 0$ which
follow also from the 
compressibility sum rule,
$\Pi(q=0) = \int d\varepsilon \left ( -{df(\epsilon)}/{d\varepsilon}
\right ) N(\varepsilon_q)$.
For small $q$, $\Pi_{\rm intra}(q)$ decreases as $1-q^2/2k_F^2$, and
$\Pi_{\rm inter}(q)$ increases as $q^2/2k_F^2$. This behavior
comes from the overlap factor $F_{ss'}$ in Eq. (\ref{pol}).
For SLG intraband (interband) polarizability decreases (increases)
linearly as q increases and these two effects exactly cancel out up to
$q=2k_F$, which gives rise to 
the total static polarizability being constant for $q<2k_F$ as in the
2DEG. However, for BLG the cancellation of two
polarizability functions is not exact
especially for $q>k_F$ because of the enhanced backscattering, so the
total polarizability increases as $q$ approaches  $2k_F$, which
means screening increases as $q$ increases.
Thus BLG, in spite of being a 2D systems, does not have a constant
Thomas-Fermi screening up to $q=2k_F$ as exists in SLG and 2DEG.

A qualitative difference between SLG and BLG polarizability functions
is at $q=2k_F$. Due to the suppression of $2k_F$ backward
scattering in SLG, the total polarizability as well as its first
derivative are continuous. In BLG, however, the large angle scattering is 
enhanced due to chirality, which gives rise to the
singular behavior of polarizability at $q=2k_F$. Even though the
BLG polarizability is continuous at $q=2k_F$,  it has a sharp cusp
and its derivative is discontinuous at $2k_F$, diverging as
$q$ approaches $2k_F$, i.e., 
as $q \rightarrow 2k_F$, $d\Pi(q)/dq \propto
1/\sqrt{q^2-4k_F^2}$. This behavior is exactly the same as that of
the regular 2DEG, which also has a cusp at $q=2k_F$ in addition to
being constant in the $0 \le q \le 2k_F$ region.
Note that in SLG this non-analytic behavior of polarizability occurs
in the second derivative, $d^2\Pi(q)/dq^2 \propto 1/\sqrt{q^2-4k_F^2}$. 

In the large momentum transfer regime, $q > 2k_F$, the BLG
polarizability 
approaches a constant value (intrinsic polarizability $\Pi^0$), i.e.,
$\Pi(q) \rightarrow  N_0 \log 4$, 
because the interband transition dominates over the intraband contribution
in the large wave vector limit. 
This is very different 
from that of a 2DEG where the static polarizability falls
off rapidly ($\sim 1/q^2$) for $q >2k_F$\cite{rmp} and SLG where the
polarizability increases linearly as $q$. 
Thus,
the large $q$ behavior of dielectric screening becomes $\epsilon(q
\rightarrow \infty) \rightarrow  1+g\pi e^2/(8\kappa v_F)$ for SLG
and $\epsilon(q \rightarrow \infty) \rightarrow 1$ for both BLG and
2DEG.

The strong cusp in BLG $\Pi(q)$ at $q=2k_F$ leads to Friedel oscillations
in contrast to the SLG behavior.
The leading oscillation term in the screened potential at large
distances from a point charge
$Ze$  can be calculated as 
\begin{equation}
\phi(r) \sim -\frac{e}{\kappa}\frac{4q_{TF}k_F^2}{(2k_F+Cq_{TF})^2}
\frac{\sin(2k_Fr)}{(2k_Fr)^2},
\end{equation}
where $C=\sqrt{5}-\log[(1+\sqrt{5})/2]$, which is similar to the 2DEG 
except for the additional constant $C$ ($C=1$ for  2DEG), but 
different from 
SLG where Friedel oscillations scale as
$\phi(r) \sim \cos(2k_Fr)/r^3$.
The enhanced singular behavior of the BLG screening function at $q=2k_F$
has 

\noindent
other interesting 
consequences related to Kohn anomaly \cite{kohn} and
RKKY interaction, which we discuss below.


The strong cusp in the BLG polarizability at $q=2k_F$, as can be
clearly seen in Fig. 1(c), indicates that the screened BLG acoustic
phonon frequency would manifest a strong Kohn anomaly, i.e. an
observable dip structure in the phonon frequency at $q=2k_F$. It is
obvious from 
Fig. 1(c), and from the discussion above based on our
analytical results for $\epsilon(q)$, that the screened phonon
dispersion will exhibit a much stronger Kohn anomaly in the BLG than
in the SLG with the 2DEG being intermediate. 
This arises from the stronger singularity at $q=2k_F$ manifesting in
the {\it first} derivative $d\Pi(q)/dq$ in the BLG rather than in the
{\it second} derivative, $d^2\Pi/dq^2$, in the SLG.
Tuning the value of $k_F$
by changing the carrier density through the applied gate voltage, it
should be possible to verify that the Kohn anomaly is indeed
associated with the $2k_F$ screening behavior in the BLG. In fact, the
BLG Kohn anomaly would be, due to the very strong $q=2k_F$ cusp in the
polarizability, rather similar to the 1D Peierls instability
\cite{Peierls} since the 
$q=2k_F$ screening behavior in BLG is qualitatively similar to the 1D
electron system \cite{Lai}.

The polarizability function in Eq. (\ref{piq}) also determines the
RKKY interaction between 
two magnetic impurities due to the induced spin density 
(here we consider magnetic impurities located at 
the interface between BLG and substrate, so they do not break any
symmetry). The RKKY interaction (or induced spin density) is  
proportional to the Fourier transform of $\Pi(q)$. The
conventional form of the exchange interaction between the localized
moment $S$ and electron-spin density $s(r)$ is given by
$V(r)=JS({\bf R})s({\bf r})\delta({\bf R-r})$,
where $J$ is the exchange coupling
constant. The RKKY interaction between two localized moments via the
conduction electrons may then be written in the following form:
$H_{RKKY}(r)= J^2S_1S_2 \Pi({\bf r})$,
where $\Pi(r)$ is the Fourier transform of the 
static polarizability $\Pi(q)$.

First consider intrinsic BLG. The Fourier transform of
Eq. (\ref{pi0b}) simply becomes a $\delta$-function because $\Pi^0(q)$ is
a constant, i.e. $\Pi(r) = N_0\log(4) \delta(r)$. This indicates that
the localized magnetic moments are not correlated by the long range
interaction and there is no induced spin density.
In SLG, the Fourier transform of polarizability
(Eq. (\ref{pi0s})) diverges (even though $\Pi(r)$ formally scales as $1/r^3$,
its magnitude does not converge), which means that intrinsic  SLG is
susceptible to ferromagnetic ordering in the presence of magnetic impurities
\cite{Voz,brey} due to the divergent RKKY coupling.

In doped (or gated) BLG the oscillatory term in RKKY interaction is
restored due to the singularity of
polarizability at $q=2k_F$, and the oscillating behavior dominates at
large $k_Fr$. At large 
distances $2k_Fr\gg 
1$, the dominant oscillating term in $\Pi(r)$ is given by                   
\begin{equation}
\Pi(r) \sim N_0 \frac{k_F^2}{2\pi}\frac{\sin(2k_Fr)}{(k_F r)^2}.
\end{equation}
This is the same RKKY interaction as in a regular 2DEG, and it
decreases as $1/r^2$, in contrast with $1/r^3$
behavior in SLG. \cite{wunsch}



In conclusion, we calculate analytically the static wave vector
dependent polarizability of both undoped and doped bilayer
graphene within RPA. 
For undoped BLG we find that screening is enhanced by a factor of
$\log4$ compared with ordinary 2D screening. The RKKY 
interaction in undoped BLG is zero-ranged ($\delta$-function)
and therefore, no spin density is induced. 
The doped BLG screening function shows  strongly 
enhanced  Kohn anomaly at $2k_F$ compared with the corresponding SLG and 
2DEG situations, which give rise to the usual RKKY interaction and
Friedel oscillation.
We show that BLG screening properties are qualitatively different from
SLG screening behavior in all wave vector regimes ($q<2k_F$, $q>2k_F$,
and $q=2k_F$) with the BLG screening having a strong cusp at $q=2k_F$.
Our theory applies only 
in the low density regime ($n<5\times 10^{12} cm^{-2}$),
where the band dispersion is
quadratic and only the lowest subband  
is occupied \cite{bilayer}.
There are obvious implications of our results for BLG carrier
transport limited by screened Coulomb scattering -- in particular, the
strong $2k_F$ anomaly in screening will lead to  strong
temperature dependence in $dc$ transport  at
low ($T<<T_F$) temperatures. This is in sharp contrast to SLG where
$2k_F$ backscattering is suppressed.

 This work is supported by U.S. ONR, NSF-NRI, and
SWAN.

\end{document}